%% file: MobilePoser.tex
\documentclass[sigconf]{acmart}
\AtBeginDocument{%
  \providecommand\BibTeX{{%
    \normalfont B\kern-0.5em{\scshape i\kern-0.25em b}\kern-0.8em\TeX}}}

\usepackage{multirow}
\usepackage{color, colortbl}
\usepackage{bm}
\usepackage{pifont}
\usepackage{url}
\usepackage{enumitem}
\usepackage{hyperref}
\usepackage{stfloats} 
\hypersetup{
    colorlinks=true,
    linkcolor=blue,
    filecolor=magenta,      
    urlcolor=magenta,
    pdftitle={Overleaf Example},
    pdfpagemode=FullScreen,
}

\copyrightyear{2024}
\acmYear{2024}
\setcopyright{rightsretained}
\acmConference[UIST '24]{The 37th Annual ACM Symposium on User Interface Software and Technology}{October 13--16, 2024}{Pittsburgh, PA, USA}
\acmBooktitle{The 37th Annual ACM Symposium on User Interface Software and Technology (UIST '24), October 13--16, 2024, Pittsburgh, PA, USA}
\acmDOI{10.1145/3654777.3676461}
\acmISBN{979-8-4007-0628-8/24/10}

\begin{document}

\title{MobilePoser: Real-Time Full-Body Pose Estimation and 3D Human Translation from IMUs in Mobile Consumer Devices}


\settopmatter{authorsperrow=2}

\author{Vasco Xu}
\affiliation{%
  \institution{University of Chicago}
  \city{Chicago}
  \country{USA}}
\email{vascoxu@uchicago.edu}

\author{Chenfeng Gao}
\affiliation{%
  \institution{Northwestern University}
  \city{Evanston}
  \country{USA}}
\email{chenfenggao2029@u.northwestern.edu}

\author{Henry Hoffmann}
\affiliation{%
  \institution{University of Chicago}
  \city{Chicago}
  \country{USA}}
\email{hankhoffmann@cs.uchicago.edu}

\author{Karan Ahuja}
\affiliation{%
  \institution{Northwestern University}
  \city{Evanston}
  \country{USA}}
\email{kahuja@northwestern.edu}

\renewcommand{\shortauthors}{Xu, et al.}
\newcommand{\vasco}[1]{{\color{cyan}(Vasco: #1)}}
\newcommand{\todo}[1]{\textcolor{red}{[TODO: #1]}}

\begin{abstract}
\input{sections/0_abstract}

\end{abstract}



\begin{teaserfigure}
    \centering
    \includegraphics[width=\textwidth]{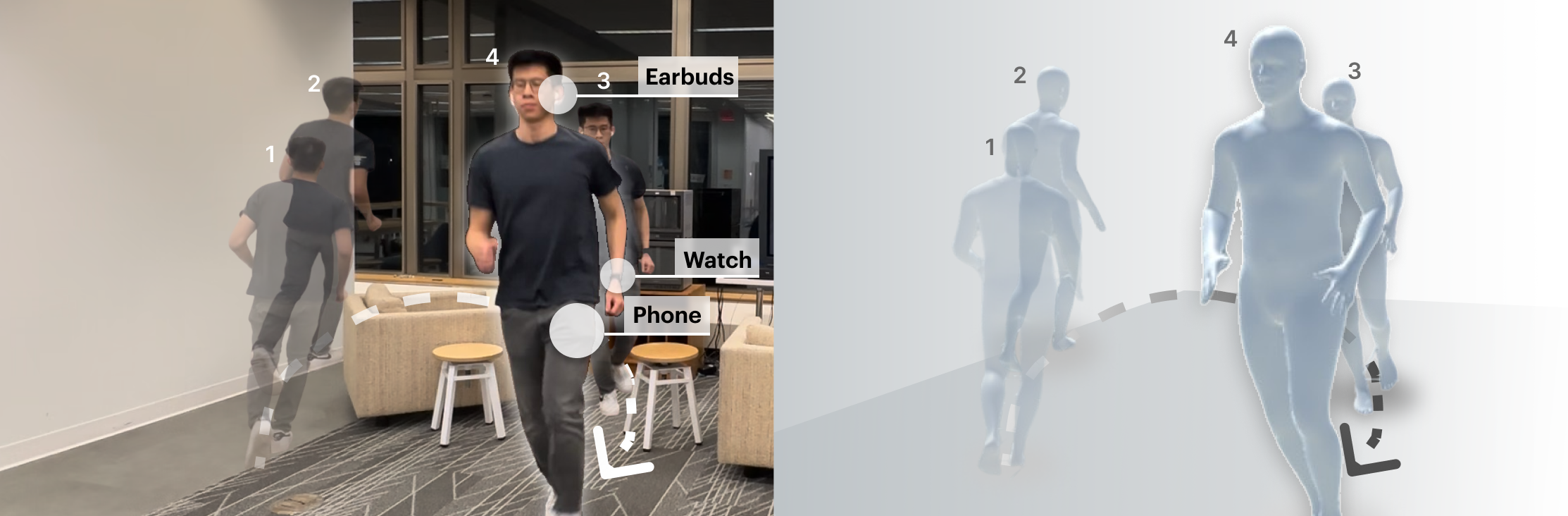}
    \caption{MobilePoser uses any subset of consumer mobile devices (phones, watches, earbuds) available to estimate full-body pose and global translation. }
    \Description{}
    \label{fig:teaser}
\end{teaserfigure}


\maketitle

\input{sections/1_introduction}

\input{sections/2_relatedwork}

\input{sections/3_mobileposer}
\input{sections/4_evaluation}

\input{sections/5_applications}
\input{sections/6_opensource}

\input{sections/7_limitations_future_work}
\input{sections/8_conclusion}

\begin{acks}
We thank Jianru Ding from the University of Chicago and Zeya Chen from the Institute of Design, Illinois Institute of Technology for helping film the video. Vasco Xu's and Henry Hoffmann's work on this project is supported by NSF (CCF-1823032 and CNS-1956180). 
\end{acks}


\bibliographystyle{ACM-Reference-Format}
\bibliography{references}

\end{document}

%% file: sections/0_abstract.tex
There has been a continued trend towards minimizing instrumentation for full-body motion capture, going from specialized rooms and equipment, to arrays of worn sensors and recently sparse inertial pose capture methods. However, as these techniques migrate towards lower-fidelity IMUs on ubiquitous commodity devices, like phones, watches, and earbuds, challenges arise including compromised online performance, temporal consistency, and loss of global translation due to sensor noise and drift. Addressing these challenges, we introduce MobilePoser, a real-time system for full-body pose and global translation estimation using any available subset of IMUs already present in these consumer devices. MobilePoser employs a multi-stage deep neural network for kinematic pose estimation followed by a physics-based motion optimizer, achieving state-of-the-art accuracy while remaining lightweight. We conclude with a series of demonstrative applications to illustrate the unique potential of MobilePoser across a variety of fields, such as health and wellness, gaming, and indoor navigation to name a few.

%% file: sections/1_introduction.tex
\begin{figure*}
    \centering
    \includegraphics[width=1\textwidth]{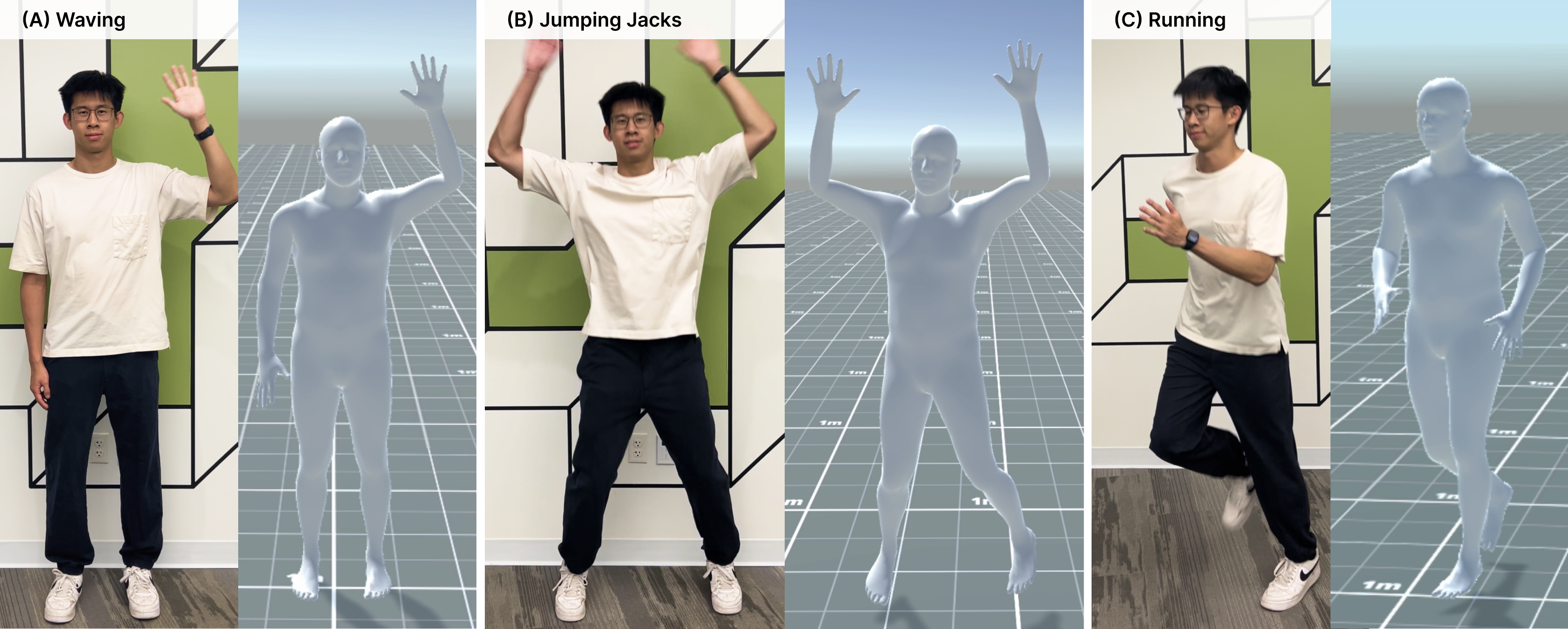}
    \caption{Real-time global pose estimation powered by MobilePoser: (A) Person with smartwatch (left wrist) waving their hands. (B) Person with smartwatch (left wrist) performing jumping jacks. (C) Person wearing a smartwatch (left wrist) and carrying a phone in their right pocket running.}
    \label{fig:video-figs}
\end{figure*}

\section{Introduction}
Full-body motion capture has numerous applications in gaming, fitness, and virtual and augmented reality (VR/AR), enabling immersive experiences and context-aware interactions. While vision-based approaches for 3D human pose estimation have shown great promise, they require subjects to be within the camera's field of view, limiting their practicability for mobile and on-the-go applications. In contrast, inertial measurement unit (IMU) based techniques offer an attractive alternative, enabling less intrusive and occlusion-free user digitization \cite{ahuja2024practicalrichuserdigitization}.

Commercial systems such as Xsens~\cite{xsens} use up to 17 special-purpose sensors to provide highly accurate pose estimations. However, such approaches are intrusive, making them undesirable for everyday use. Consequently, there has been a trend towards minimizing instrumentation. Sparse inertial pose capture methods, such as TransPose~\cite{yi2021transpose} and DIP~\cite{huang2018deep}, use 6 IMUs to achieve a balance between accuracy and practicality. Yet, these methods still require expensive and special-purpose IMUs attached to specific body joints. To enable full-body motion tracking without any external infrastructure, IMUPoser~\cite{mollyn2023imuposer} leverages IMUs in devices we already carry around with us, namely smartphones, smartwatches, and earbuds. These commodity devices, however, use lower-fidelity IMUs, which compromises online performance, temporal consistency, and global translation estimation.

In this work, we present MobilePoser, a real-time user digitization technique that tracks both \textit{poses} and global \textit{movement} (referred to as \textit{translation}) using consumer devices (Figure~\ref{fig:teaser}) such as watches, phones and earbuds. To enable on-the-go motion tracking without any external infrastructure, we must address a set of unique challenges. First, the number of instrumented points is dynamically changing and sparse (at most three devices, with as few as one) \footnote{Note, we count the left and right earbuds as a unified single IMU stream}, making the problem highly under-constrained. Second, IMUs do not directly measure positional data, making global translation tracking non-trivial. Additionally, noise and drift from the low-cost IMUs found in commodity devices complicates pose and translation estimation. Finally, such a system should operate directly on-device for real-time use, anywhere, anytime.

MobilePoser tackles these challenges by employing a multi-stage approach. For pose estimation, it utilizes a deep neural network (DNN) to predict full-body pose from the available IMU data, followed by a physics-based optimization step to ensure spatio-temporal consistency and plausible kinematics. This greatly helps resolve ambiguous instrumented joint motion profiles, such as differentiating between waving (Figure~\ref{fig:video-figs} A) versus jumping jacks (Figure~\ref{fig:video-figs} B) from only a single smartwatch on the wrist. To aid in generalizability, the model is trained on a large dataset of synthesized IMU measurements generated from high-quality motion capture (MoCap) data. For global translation estimation, MobilePoser employs a hybrid approach that fuses predictions from a foot contact-based method and a DNN-based method that directly regresses the root joint velocity. This combination enables accurate and robust translation estimation, even in challenging scenarios where both feet are in motion together (Figure~\ref{fig:video-figs} C). Importantly, MobilePoser is optimized to run on-device, achieving real-time performance of 60 frames per second on a smartphone (iPhone 15 Pro), making it suitable for mobile applications.


In summary, MobilePoser makes the following key contributions:
\begin{enumerate}
    \item It presents a novel framework for inertial translation estimation using consumer devices, enabling accurate tracking of global movement without specialized hardware.
    \item It achieves state-of-the-art full-body pose estimation across various on-body configurations of commodity IMU devices, demonstrating robust performance with as few as one and up to three wearable devices.
    \item It provides an open-source implementation that runs in real-time on edge devices, making it accessible and practical for widespread use.
\end{enumerate}

%% file: sections/2_relatedwork.tex
\section{Related Work}

\begin{table*}

\centering
\begin{tabular}{lcccccccc}
\hline
System & \# Inst. Joints & FPS & Consumer Device  & Translation &  MPJVE (cm) & Jitter ($10^{2}m/s^3$) \\
\hline
Xsens \cite{xsens}  & 17 & 120 & $\times$ & \checkmark & - & - \\
SIP \cite{von2017sparse}  & 6 & 60 & $\times$ & \checkmark & 7.7 & 3.8 \\
DIP \cite{huang2018deep} & 6 & 29 & $\times$ & $\times$ & 8.9 & 30.13 \\
TransPose \cite{yi2021transpose}  & 6 & 90 & $\times$ & \checkmark & 7.1 & 1.4 \\
PIP \cite{yi2022physical} & 6 & 60 & $\times$ & \checkmark & 5.9 & 0.24 \\
IMUPoser \cite{mollyn2023imuposer} & 1--3 & 25 & \checkmark & $\times$ & 12.1 & 1.9 \\
\rowcolor{blue!10} MobilePoser (our work) & 1--3 & 60 & \checkmark & \checkmark & 10.6 & 0.97 \\
\hline
\end{tabular}
\caption{Comparison with key prior work on the DIP-IMU dataset.}
\label{tab:dip-key-comparison}
\end{table*}

\subsection{User Digitization with External Sensors}

Commercial motion capture systems such as OptiTrack~\cite{optitrack} and Vicon~\cite{vicon} use specialized hardware, such as multiple calibrated high-speed infrared cameras, to track retroreflective markers attached to a user's body. Such setups are commonly used in games, movies and character animations that require millimeter accuracy and are the gold standard of motion capture. The expensive infrastructure required by commercial systems, makes them impractical for everyday use. Therefore, much research has been devoted to instrumentation-free approaches using monocular cameras. Such approaches generally rely on RGB~\cite{bogo2016keep, goel2023humans, rajasegaran2021tracking} or depth~\cite{kinect} cameras based computer vision techniques to predict body pose. 

There also exists specialized external hardware for pose tracking in Extended Reality (XR). For example, the HTC Vive~\cite{vive}, PlayStation VR~\cite{playstationvr} and Oculus Rift~\cite{parger2018human} track the head, handheld controllers and other limb-borne accessories using external sensor base stations for Virtual Reality (VR) applications. The un-sensed joints are estimated with inverse kinematics~\cite{jiang2016real} or learning-based methods \cite{jiang2022avatarposer,ponton2023sparseposer}. Other non-optical external approaches for pose estimation include capacitive sensing~\cite{zhang2018wall}, magnetic fields~\cite{trakstar, polhemus}, RF~\cite{zhao2018through}, and mechanical linkages~\cite{sutherland1968head}.

\subsection{User Digitization with non-IMU Worn Sensors}

Wearable sensors provide a portable and flexible alternative to external sensors. For example, MI-Poser~\cite{arakawa2023mi} uses magnetic tracking in wristbands and AR glasses to estimate upper-body poses. Other works have explored wrist-worn cameras~\cite{wu2020back, kim2012digits}, EMG sensors~\cite{liu2021neuropose}, EIT sensors \cite{kyu2024eitpose}, wrist-worn antennas~\cite{kim2022etherpose} and depth sensor armbands~\cite{devrio2022discoband}. However, these works focus solely on capturing the motion of specific body parts (e.g., wrist or upper-body). 

To capture full-body motion, a popular approach is to use body-mounted cameras coupled with computer vision techniques~\cite{shiratori2011motion, ahuja2022controllerpose}. Other works have explored different sensor technologies such as ultrasonic sensors~\cite{vlasic2007practical} and RFID ~\cite{jin2018towards}. Nevertheless, these works require users to wear sensors they do not already have. Pose-On-The-Go~\cite{ahuja2021pose} addresses this by estimating full-body pose via extreme sensor fusion, leveraging a phone's front and rear cameras, thus requiring no special instrumentation. However, its computationally expensive and relies heavily on heuristics to power body poses, often resulting in unnatural motions. MobilePoser differentiates itself by focusing on full-body pose estimation using power-efficient IMUs already found in consumer devices, such as smartphones, smartwatches, and earbuds. 


\subsection{User Digitization with IMU Worn Sensors}

Commercial motion capture systems, such as Xsens~\cite{xsens}, use a large number of inertial sensors (typically 17) strapped to the body to provide high-quality motion capture. These setups consist of homogeneous, high-grade IMUs that are calibrated for noise and have known positions on the body, resulting in a less ill-posed problem compared to using sparse, heterogeneous sensors. However, such an approach is highly inconvenient and intrusive for everyday use.

To address this limitation, researchers have explored reconstructing human motions from a reduced number of sensors. Works such as SIP~\cite{von2017sparse}, DIP~\cite{huang2018deep}, PIP~\cite{yi2022physical}, TIP~\cite{jiang2022transformer}, and  TransPose~\cite{yi2021transpose} have demonstrated the feasibility of using only 6 commercial-grade Xsens IMU sensors for full-body motion capture. Works have further explored integrating other input modalities (e.g. UWB~\cite{armani2024ultra} and egocentric images~\cite{yi2023egolocate}) in addition to the 6 IMUs for increased performance. All these approaches leverage the homogeneity and known calibrated positions of the sensors to achieve accurate pose estimation. However, even 6 sensors can be cumbersome for on-the-go applications, especially those that require passive sensing.

Recent research has investigated even sparser IMU configurations using commodity devices. IMUPoser~\cite{mollyn2023imuposer}, which is  most closely related to our work, performs pose estimation using any combination of smartphone, smartwatch, and earbuds. While IMUPoser tackles the challenges of heterogeneous sensor quality for pose estimation, it lacks global translation due to IMU noise and drift, and contains unrealistic spatio-temporal motion artifacts. Additionally, IMUPoser runs on a laptop at 25Hz, limiting its practicality for real-time mobile applications.

In contrast, MobilePoser addresses these limitations by demonstrating improved pose estimation accuracy on widely used benchmarks \textit{while also estimating global translation} (see Table~\ref{tab:dip-key-comparison}). Furthermore, our system is designed to run fully on-device, achieving real-time performance of 60 fps on edge mobile devices. This enables MobilePoser to provide a more practical and accessible solution for on-the-go motion capture using commodity devices.






%% file: sections/3_mobileposer.tex
\begin{figure*}
    \centering
    \includegraphics[width=\textwidth]{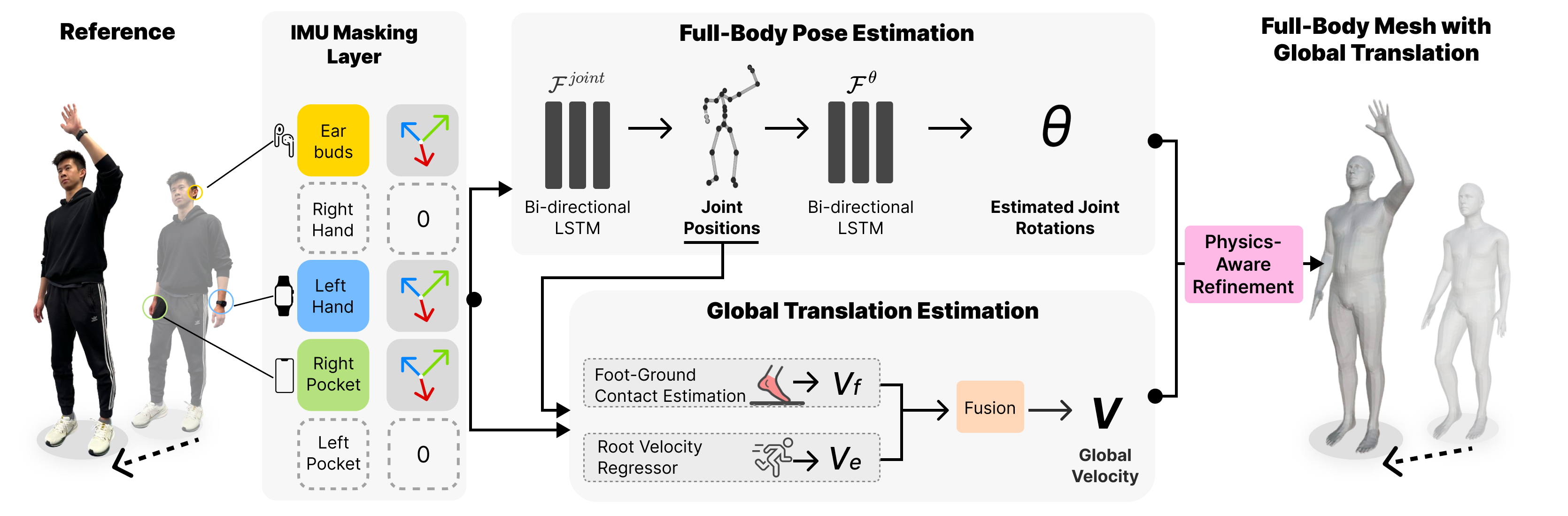}
    \caption{MobilePoser system overview. MobilePoser accepts any available subset of IMU data from the user and masks absent devices by setting their values to zero. The IMU data is then fed into two main modules: (1) Pose Estimation, which first estimates joint positions followed by joint rotations, and (2) Translation Estimation, which combines foot-ground contact probabilities with a direct neural network-based approach to regress global velocity. Finally, a Physics Optimizer refines the predicted joint rotations and global translation to ensure they satisfy physical constraints.}
    \label{fig:system-overview}
\end{figure*}

\section{MobilePoser}

Estimating a user's full-body pose from a sparse set of IMU observations is a severely under-constrained problem as it aims to infer a high-dimensional quantity, i.e., the full-body pose, from low-dimensional observations that only capture partial motion at each instrumented point. Moreover, multiple possible solutions could explain the observed data, making it challenging to determine the correct pose. To tackle these challenges, we introduce MobilePoser, a system that leverages data-driven learning and physics-based optimization to estimate accurate and plausible full-body poses and global translations from sparse IMU inputs. Figure~\ref{fig:system-overview} provides an overview of our pipeline, which we describe in detail in the following sections.

\subsection{System Input}
\label{sec:device-combos}

MobilePoser takes as input acceleration and orientation readings from IMUs across any subset of three consumer devices: smartphones, smartwatches, and earbuds. Each of these devices can be placed at different body locations, resulting in various possible combinations. For instance, a smartphone can be stored in the left or right pocket, held in the left or right hand, placed next to the head during a call, or not carried by the user at all. Similarly, smartwatches can be worn on either wrist or not worn at all, while earbuds can be worn, placed in a charging case stored in either pocket, or not carried by the user.

Following IMUPoser~\cite{mollyn2023imuposer}, we consider 24 plausible device-location combinations across five body locations: right pocket, left pocket, right wrist, left wrist, and head. These combinations cover the various ways users might carry or wear their devices throughout the day. Regardless of the input device combination, our model expects IMU data from the five predefined body locations. 

The IMU signal at each location consists of acceleration (3 values) and orientation (a $3 \times 3$ rotation matrix), resulting in a total of 12 IMU values per location. Across all five locations, this yields an input vector $\bm{x} \in \mathbb{R}^{60}$.  However, since at any given time only a subset of 1--3 devices may be present, data from absent devices is masked and set to zero. This masking approach allows us to build a unified model that can handle the varying number of available devices and their changing on-body location seamlessly. This further eliminates the need for training separate models for each possible combination, making the system more practical and efficient. 


\subsection{Full-Body Pose Estimation}

To learn a mapping from IMU input to full-body pose, we employ a data-driven, multi-stage neural network approach. Specifically, our pose estimation network consists of two submodules: Joint predictor ($\mathcal{F}^{joint}$) and Rotation predictor ($\mathcal{F}^{\theta}$). More specifically, $\mathcal{F}^{joint}$ estimates joint positions as an intermediate task and $\mathcal{F}^{\theta}$ solves for the joint angle orientations. Both submodules use a bidirectional LSTM (bi-LSTM), to model both spatial and temporal information~\cite{huang2018deep}. We input data into both submodules in a sliding-window fashion with window length $N$. 

\subsubsection{Joint Pose Estimation ($\mathcal{F}^{joint}$)}
This module estimates the joint positions from a sequence of IMU measurements. We explicitly estimate joint positions as an intermediate step, as it helps extract useful information from linear accelerations due to its linear correlation with joint positions~\cite{yi2021transpose}. The input to $\mathcal{F}^{joint}$ is $\bm{x}^{imu}(t) = \left[ \bm{x}_{t-N}, \ldots, \bm{x}_{t} \right]$, where $t$ is the current time step and $N$ is the time window length. The output are the root (pelvis) relative 3D positions of the 24 SMPL body joints \cite{SMPL:2015} $\bm{p}(t) = \left[ \bm{p}_{t-N}, \ldots, \bm{p}_{t} \right] \in \mathbb{R}^{N \times 72}$. The loss function used to train this network is: 
\begin{equation}
\mathcal{L}_{joint} = \left\| \mathbf{p} - \mathbf{p}_{GT} \right\|^{2}_{2}
\end{equation}
where the subscript $GT$ denotes the ground truth and $p$ represents the full-body SMPL joint positions.  


\subsubsection{Joint Rotation and Body Mesh Estimation ($\mathcal{F}^{\theta}$)}
Here we employ a neural kinematic estimator to regress joint rotations from the previously estimated positions. We concatenate the joint coordinates from $\mathcal{F}^{joint}$ with IMU measurements, which serves as the input to $\mathcal{F}^{\theta}$.  Note, while the SMPL body encodes 24 joints, only 18 are relevant from a rotation prediction perspective as the fingers, wrist and toes are independent of the on-body IMUs and are hence set to identity rotation matrices \cite{yi2021transpose}. The outputs of the network are the 18 root relative joint orientations represented as 6D rotations: $\bm{\theta}(t) = \left[ \bm{\theta}_{t-N}, \ldots, \bm{\theta}_{t} \right] \in \mathbb{R}^{N \times 108}$.

Our joint rotation loss consists of three terms: $\mathcal{L}_{ori}$, $\mathcal{L}_{pos}$, $\mathcal{L}_{jerk}$. The loss term $\mathcal{L}_{ori}$ is a standard L2 loss from the ground truth joint rotations. The term $\mathcal{L}_{pos}$ penalizes error accumulating along the kinematic chain. Finally, $\mathcal{L}_{jerk}$ promotes temporally smooth predictions, where $jerk(\theta)= \theta_{t-3} + 3\theta_{t-2} - 3\theta_{t-1} + \theta_{t}$ is a function that computes the jerk of a signal $\theta$ at time step $t$, penalizing the deviation between neighboring frames \cite{yi2021transpose}. 

Our combined joint rotation loss function can be represented as,
\begin{align}
\mathcal{L}_{\theta} &= \mathcal{L}_{ori} + \mathcal{L}_{pos} + \lambda \mathcal{L}_{jerk} \\
\mathcal{L}_{ori}  &= \left\| \mathbf{\theta} - \mathbf{\theta}_{GT} \right\|^{2}_{2} \\ 
\mathcal{L}_{pos}  &= \left\| \text{FK}(\mathbf{\theta}) - \mathbf{p}_{GT} \right\|^{2}_{2} \\ 
\mathcal{L}_{jerk} &= \sum_{t}^{T} jerk\left( \mathbf{\theta} \right)
\end{align}
where $FK(\cdot)$ is the forward kinematics function, that computes joint coordinates from joint rotations. Given the joint rotations, the parametric SMPL body model generates a corresponding body mesh with 6890 vertices.

 
\subsection{Global Translation Estimation}

Translation estimation from IMUs is challenging as they lack direct distance measurements. Moreover, IMUs are prone to noise and biases, which causes techniques such as double-integration of acceleration to rapidly accumulate errors~\cite{yan2018ridi}. Therefore, inspired by prior work~\cite{yi2021transpose, yi2022physical, lee2024mocap}, we estimate per-frame velocity of the root joint using two submodules:  a foot-ground contact ($v_f$) and a neural network based root velocity estimator ($v_e$). We fuse the output of the two submodules to obtain a final estimate of global translation. 

\subsubsection{Foot-Ground Contact based Root Velocity ($v_f$)}

Here we estimate the probability of each foot contacting the ground independently using a bi-LSTM network. The input to the model is the concatenated vector of joint positions and IMU measurements. The output of the network is the likelihood that each foot is contacting the ground, denoted as $\bm{c}_{foot} = \left[\bm{c}_{lfoot}, \bm{c}_{rfoot} \right] \in \mathbb{R}^{2}$. The foot with the higher foot-ground contact probability is defined as the supporting foot, $s = \max\{c_{\text{lfoot}}, c_{\text{rfoot}}\}$. The root velocity, $v_f(t) \in \mathbb{R}^{3}$, is then computed as the coordinate difference of the supporting foot between consecutive frames. This approach helps capture natural body motions, as movement is significantly influenced by the supporting foot's dynamics~\cite{roy2014smartphone}. For example, when walking, the body's movement is propelled forward and stabilized by the foot contacting the ground. The network is trained using binary cross-entropy loss.




\subsubsection{Neural Network based Root Velocity ($v_e$)} 
While the supporting foot contact based method yields plausible human movement, it inherently fails when both feet are not contacting the ground (e.g., when running or jumping). To accommodate such cases, we estimate per-frame root velocity directly using a neural network. We again use the predicted joint coordinates and IMU measurements as input. Compared to previous submodules that use a bi-LSTM for prediction, this module uses a unidirectional LSTM due to its capacity to capture longer historical context. The output is per-frame root velocity, denoted as $v_e(t) \in \mathbb{R}^{3}$. The network is trained using a cumulative L2 loss~\cite{yi2021transpose}. 



\subsubsection{Module Fusion.} 
Both modules offer different trade-offs in terms of predicting translation. Supporting foot provides more realistic estimates by leveraging human kinematics but fails when both feet are off the ground. On the other hand, directly estimating root velocity is more general but is highly prone to unnatural movements such as foot sliding~\cite{zhuo2023towards}. To achieve the benefits of both, we adopt the heuristic-based fusion approach, inspired by TransPose~\cite{yi2021transpose}. In summary, when the foot contact $c$ is higher than an upper-threshold $\overline{q}$, we are confident of ground contact by a foot and hence we rely on ($v_f$) for translation estimation. When the foot contact is below a lower-threshold, $\underline{q}$, we rely on ($v_e$). For intermediate probabilities, we fuse both velocity estimations using a weighted sum, to output the final global velocity estimate $v$: 
\begin{equation}
v = \dfrac{q - \overline{q}}{\underline{q} - \overline{q}} v_e + \dfrac{q - \underline{q}}{\overline{q} - \underline{q}} v_f 
\end{equation}
Following previous work~\cite{yi2021transpose}, we use $\underline{q} = 0.5$ and $\overline{q} = 0.9$.

\subsection{Physics-Aware Refinement}
\begin{figure}[b]
    \centering
    \includegraphics[width=\columnwidth]{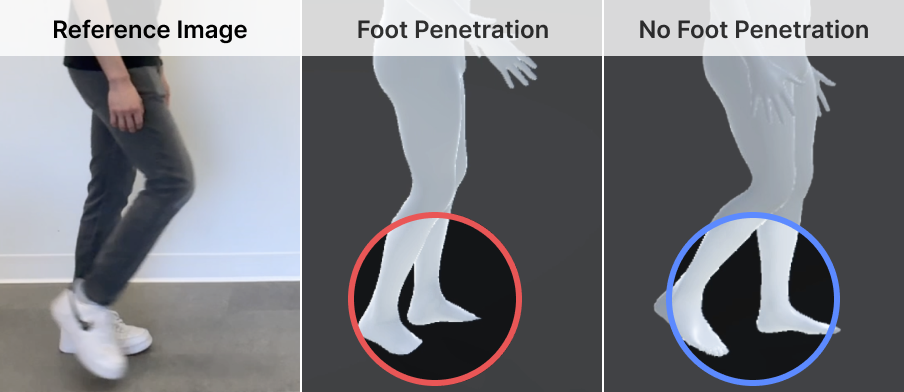}
    \caption{Demonstration of the physics optimizer's ability to reduce foot-ground penetration.}
    \label{fig:physics}
\end{figure}

Our pose and translation estimation networks output the user's global pose based on a history of IMU measurements. When trained on sufficiently large amounts of data, the full-body pose estimation and global translation estimation neural networks learn the human motion manifold and produce realistic poses. However, despite the best modeling efforts, the outputs may still contain inter-mesh penetration, temporal artifacts such as jitter, foot-floor penetration and foot skating. To address these issues, we add an off-the-shelf physics motion optimizer~\cite{yi2022physical}. The physics optimizer uses two proportional derivative (PD) controllers to compute the desired acceleration of the simulated character that best reproduces the estimated pose while satisfying physical constraints, such as the equation of motion~\cite{featherstone2014rigid}. The inputs to the physics optimizer are the estimated joint angles $\theta$, the foot-ground contact probabilities $c_{foot}$, and the neural network based root velocity $v_{e}$. The outputs are the optimized joint angles and global translation with reduced jitter and foot-ground penetration (Figure~\ref{fig:physics}). For a detailed overview of the physics optimizer, we refer readers to PIP~\cite{yi2022physical}. 

\subsection{Real-time Inference}

We implement proof-of-concept applications in iOS, using an Apple iPhone 15 Pro, Apple Watch Series 9 and Apple AirPods Pro. The iPhone, Apple Watch and AirPods sample IMU data at 60, 60 and 25 Hz respectively. For uniformity, we convert all the IMU data to 60 Hz by upsampling the AirPods.  

We employ the active device selection strategy proposed by IMUPoser~\cite{mollyn2023imuposer}, wherein the UWB and inertial data is used to track the active devices and their on-body locations. For initial prototyping, the Apple Watch and AirPods communicate over Bluetooth to the iPhone, which streams data to a MacBook Air 2022 via socket. Post connection, a small calibration step is performed to align the IMU measurements with the training data, similar to prior work~\cite{mollyn2023imuposer, yi2021transpose, huang2018deep}. Following the setup, data is streamed to the laptop for pre-processing, inference and then relayed to Unity applications for visualization. 


To further prototype an on-device edge model, we convert our trained PyTorch model into CoreML with mixed precision quantization and evaluate its performance. On an iPhone 15 Pro, our model incurs \textasciitilde 14ms model inference time running at 60 Hz, capped by input IMU sampling rate. 



\section{Data Synthesis and Model Training}

Model training requires a large collection of synchronized IMU measurements and corresponding SMPL body poses. We leverage the AMASS~\cite{mahmood2019amass} MoCap dataset, which provides an extensive collection of such data(\textasciitilde40 hours), including translation. 


\subsection{Full-Body Pose Estimation}

Our models expect IMU measurements as input. We synthesize IMU data following the approach proposed in DIP~\cite{huang2018deep}. In summary, we place \textit{virtual} sensors on the corresponding SMPL mesh vertices (left and right wrists, left and right pockets, and the head) and obtain joint rotations via limb orientations, while acceleration values are computed using finite differences. During training, we scale down the acceleration by a factor of 30 $m/s^2$, such that its values are on a similar scale to orientations, for better learning. Of note, we do not normalize our IMU measurements to a root joint (e.g., the pelvis), as the number of available devices can vary. 

\subsection{Global Translation Estimation}

The translation estimation networks require (1) binary labels for foot-ground contact states and (2) per-frame root velocity values. To generate foot-ground contact states, we assume that a foot in contact with the ground displays very little movement between frames. Therefore, when the movement of one foot between consecutive frames is less than a threshold $u$, then we consider it to be contacting the ground. We set $u = 0.008$, following previous work~\cite{yi2021transpose}. To train $v_e$, we require per-frame root velocities. Since the AMASS dataset provides root position data, we can compute root velocities as the coordinate difference of the root position between consecutive frames.

\subsection{Training Setup and Procedure} 
We train our models on a NVIDIA A40 GPU, which takes roughly a day for all modules and device-combinations. In total, our model has \textasciitilde 6.7M trainable parameters. Each module is trained separately using a batch size of 256 and the Adam optimizer~\cite{kingma2014adam} with a learning rate of lr = $10^{-3}$ for 80 epochs. We also apply a gradient clipping with norm of 1, to prevent the gradients from exploding. 

During training of $\mathcal{F}^{\theta}$, $v_e$, and $v_f$, we add Gaussian noise with $\sigma = 0.04$ to the joint positions to prevent overfitting and deal with prediction errors from $\mathcal{F}^{joint}$. We empirically set $\lambda = 10^{-5}$ when training $\mathcal{F}^{\theta}$,
to encourage temporally smooth predictions.

%% file: sections/4_evaluation.tex
\section{Evaluation}

We systematically isolate and analyze the efficacy of MobilePoser across different datasets, evaluation metrics and protocols. We show both qualitative and quantitative results, and also run ablation studies to evaluate our translation estimation design choices. 

\begin{table}[b]
\centering
\begin{tabular}{ l | c c c }
Dataset & Capture Device & Translation & Data FPS \\
\hline
DIP-IMU & Commercial & $\times$ & 60 Hz \\
TotalCapture &  Commercial & \checkmark & 60 Hz \\
IMUPoser & Consumer & \checkmark & 25 Hz \\
\end{tabular}
\caption{Real-world IMU datasets for MobilePoser Evaluation.}
\label{tab:table_datasets}
\end{table}

\subsection{Datasets}
\label{sec:IMU_datasets}


We evaluate MobilePoser on three real-world, inertial datasets, summarized in Table~\ref{tab:table_datasets}:
\begin{itemize}
    \item \textit{DIP-IMU~\cite{huang2018deep}} contains data from 10 participants, collected using commercial-grade Xsens~\cite{xsens} IMUs at 60 Hz. It includes a rich variety of activities such as arm raises, stretches, lunges, squats, and punches. However, DIP-IMU does not contain global translation data. 
    
    \item \textit{TotalCapture~\cite{trumble2017total}} provides real IMU measurements with ground-truth pose and translation, captured using commercial Xsens IMUs at 60 Hz. Following PIP~\cite{yi2022physical}, we re-calibrate the acceleration measurements to account for constant bias. 
    
    \item \textit{IMUPoser~\cite{mollyn2023imuposer}} is collected from 10 participants using consumer-grade devices: an iPhone 11 Pro, Apple Watch Series 6, and AirPods, at 25 Hz. It provides ground-truth pose and global translation data. 
\end{itemize}



\begin{figure}
    \centering
    \includegraphics[width=0.9\columnwidth]{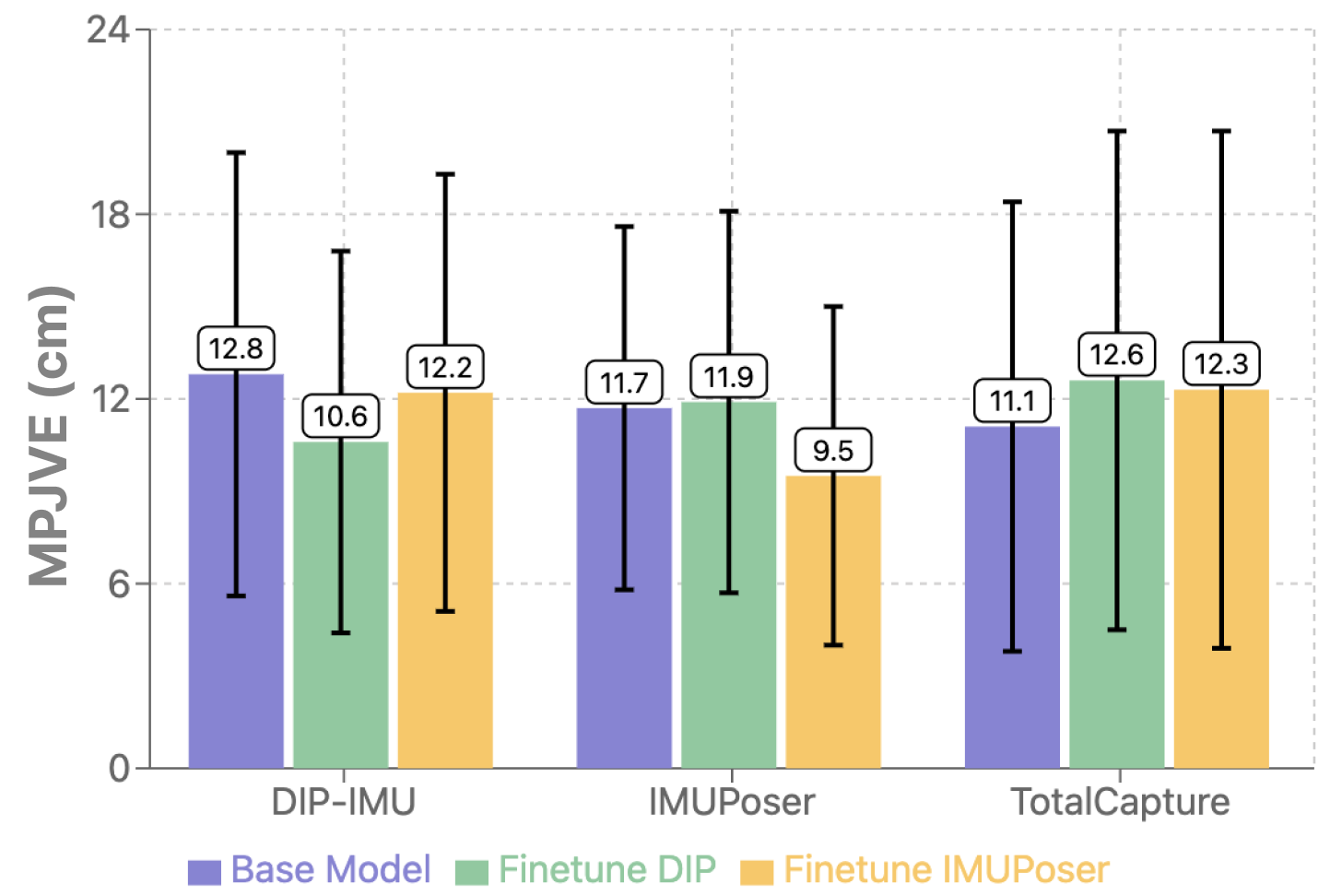}
    \caption{Comparison of MobilePoser's Full-Body Pose Estimation Error across different Evaluation Protocols on the DIP-IMU, IMUPoser and TotalCapture dataset respectively.}
    \label{fig:protocol-comparison}
\end{figure}



\subsection{Full-Body Pose Estimation}

\subsubsection{Evaluation Metrics} 

Like prior work, we use the following evaluation metrics for pose estimation (lower is better for all):

\begin{itemize}
    \item \textit{Mean Per Joint Rotation Error (MPJRE):} Measure of mean angular error across all root aligned joints in degrees (\textdegree).
    
    \item \textit{Mean Per Joint Position Error (MPJPE):} Measure of mean Euclidean distance error across all root aligned joints in centimeters (cm).
    
    \item \textit{Mean Per Joint Vertex Error (MPJVE):} Measure of mean Euclidean distance error across all root aligned vertices of the SMPL body mesh in centimeters (cm). 
    
    \item \textit{Mean Per Joint Jitter (Jitter):} Measure of mean jerk across all body joints of the predicted motion in $m/s^3$. 
\end{itemize}

We use MPJVE as our primary metric of evaluation for ease of comparison with prior work~\cite{mollyn2023imuposer}. 

\subsubsection{Evaluation Protocol} 
\label{subsubsec:eval_protocol}

We outline three evaluation protocols for training and fine-tuning to evaluate MobilePoser's efficacy across different data sources and noise profiles.

\begin{itemize}
    \item \textbf{Base Model}: We train our model on the synthetic data generated on the AMASS dataset.
    \item \textbf{Finetune DIP-IMU}: Like prior work, we train on AMASS and then fine-tune on 8 DIP-IMU participants. The 2 holdout participants are used for testing the Finetune DIP-IMU model on the DIP-IMU dataset.
    \item \textbf{Finetune IMUPoser}: We train on AMASS and fine-tune on the first 8 IMUPoser participants. The 2 holdout participants are used for testing the Finetune IMUPoser model on the IMUPoser dataset. 
\end{itemize}


\subsubsection{Accuracy across Datasets} 
Figure~\ref{fig:protocol-comparison} shows our full-body pose estimation accuracy for all three protocols across the three datasets listed in Section~\ref{sec:IMU_datasets}. Averaged across all three datasets, the MPJVE for the Base Model, Finetune DIP-IMU and Finetune IMUPoser protocols are 11.89, 11.73 and 11.33 cm respectively. It is interesting to note that the addition of commercial-grade IMU data (Finetune DIP-IMU) only improves accuracy by 1.3\% over the base model, while the addition of noisy IMU data from consumer devices (Finetune IMUPoser) results in a bigger improvement of 4.7\%. 

\subsubsection{Accuracy across Activities}
We further analyze results on different activities on the IMUPoser dataset, as it provides activity label meta-data. MobilePoser’s accuracy generalizes across most everyday activity contexts: the error (MPJVE) for locomotion is 8.2 cm (walking 7.6 cm, jogging 8.8 cm), exercises is 10 cm (kicking: 7.5 cm, jumping jacks: 11.1 cm, boxing: 11.5 cm), sitting is 11.5 cm and freestyle motions such as tennis and basketball are 9.1 cm and 11.7 cm respectively. The accuracy degrades for postures with the user lying/facing down, e.g. push-ups have higher error of 16.1 cm. 


\begin{table}
\centering
\begin{tabular}{lcccc}
\hline
System & \# Inst. Joints & MPJRE & MPJVE & Jitter \\
\hline
DIP & 6 & 17.2\textdegree & 11.2 & 3.62 \\
TransPose & 6 & 12.8\textdegree & 7.4 & 0.95 \\
PIP & 6 & 12.1\textdegree & 6.5 & 0.20 \\
IMUPoser & 1--3 & 25.6\textdegree & 15.4 & 1.30 \\
\rowcolor{blue!10} MobilePoser & 1--3 & 23.7\textdegree & 12.6 & 0.55 \\
\hline
\end{tabular}
\caption{Comparison with key prior work on the TotalCapture dataset.}
\label{tab:totalcapture-key-comparison}
\end{table}

\begin{figure}[b]
    \centering
    \includegraphics[width=0.85\columnwidth]{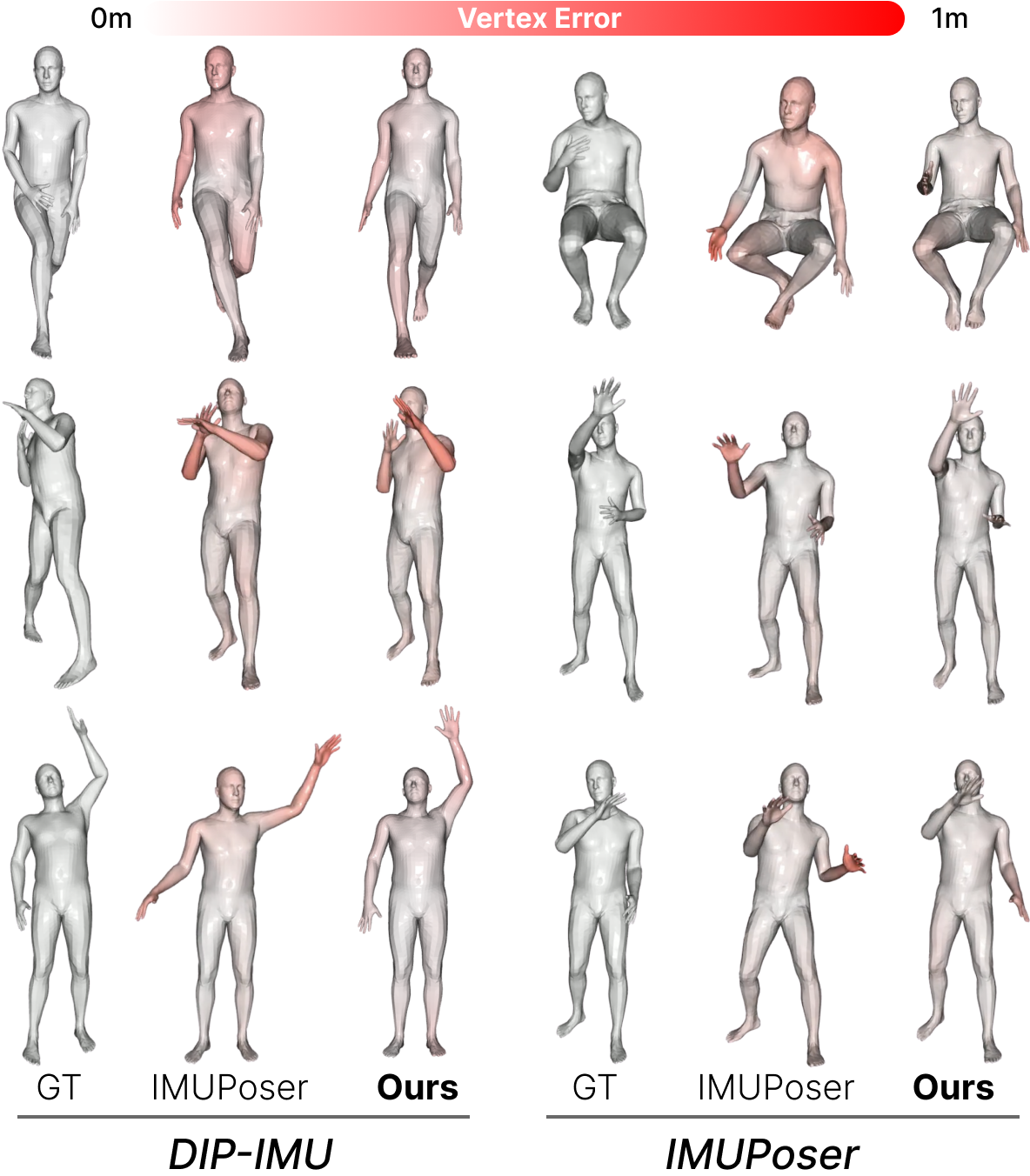}
    \caption{Qualitative comparisons between our method and IMUPoser on the DIP-IMU and IMUPoser dataset.}
    \label{fig:dip-qual}
\end{figure}

\subsubsection{Comparison with prior work}

To aid in direct comparison with prior work~\cite{mollyn2023imuposer, yi2021transpose, yi2022physical, huang2018deep}, we now make use of the Finetune DIP-IMU evaluation protocol, that is training a base model on the synthetic IMU data from AMASS and fine-tuning it on the 8 participants from DIP-IMU dataset. Tables~\ref{tab:dip-key-comparison} and ~\ref{tab:totalcapture-key-comparison} offer a quantitative comparison against key prior work, evaluated on the DIP-IMU and TotalCapture, dataset respectively. Given that our system targets a very sparse configuration of IMUs (1-3), it is unsurprising that we perform worse than systems utilizing 6 IMUs, strategically placed around the body. On the DIP-IMU and TotalCapture dataset, compared to IMUPoser, which considers the same device-location combinations, we perform significantly better displaying a 12.4\% and 18.2\% decrease in vertex error respectively. 

\begin{figure*}
    \centering
    \includegraphics[width=\textwidth]{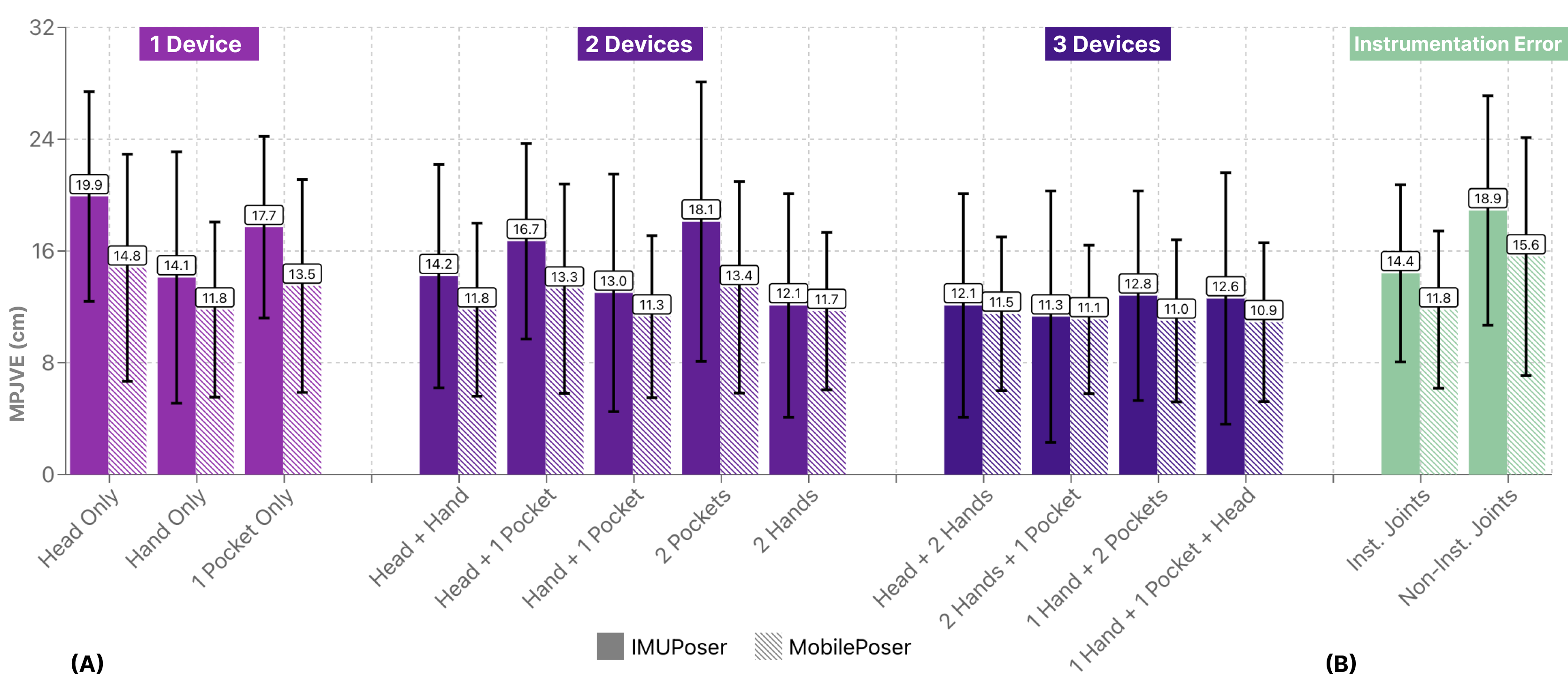}
    \caption{MPJVE comparison between IMUPoser and MobilePoser (our system) on the IMUPoser Dataset for: (A) Different on-body device combinations (B) Instrumented vs Non Instrumented joints.}
    \label{fig:device-combos}
\end{figure*}

On the IMUPoser dataset, Figure~\ref{fig:device-combos} (A) provides a detailed breakdown of accuracy for different on-body device locations. Averaging across the 1, 2 and 3 device conditions, MobilePoser outperforms IMUPoser by 24.1\%, 14.2\% and 8.7\% respectively. Furthermore, Figure~\ref{fig:device-combos} (B) provides an accuracy breakdown for the instrumented and non-instrumented joints in comparison with IMUPoser. If a limb has an IMU placed on any part, we consider all the joints pertaining to it as instrumented joints, while the rest are marked as non-instrumented. MobilePoser is 18.1\% and 17.4\% better than IMUPoser for predicting instrumented and non-instrumented joints respectively. This can be seen in Figure~\ref{fig:dip-qual} which depicts a visual comparison of our pose estimation with IMUPoser.

\subsection{Global Translation Estimation}

\begin{figure}[b]
    \centering
    \includegraphics[width=\columnwidth]{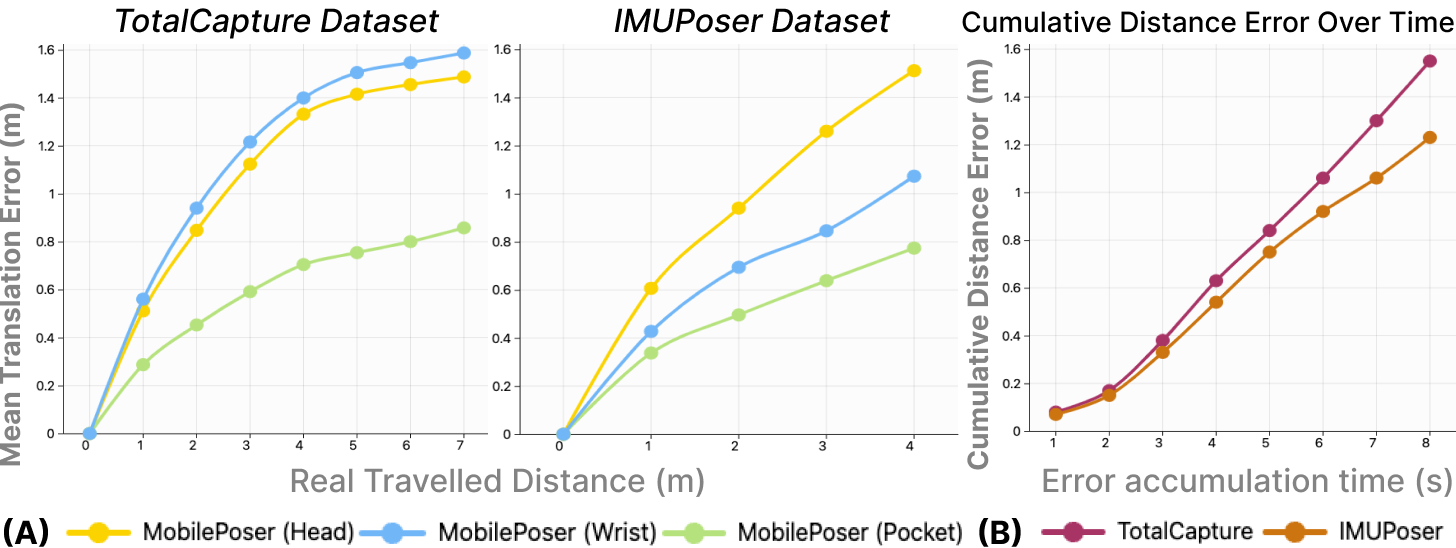}
   \caption{(A) Comparison of cumulative translation error for different instrumented joints on the IMUPoser and TotalCapture dataset. (B) Evaluation of cumulative distance errors with respect to time.}     \label{fig:translation-error}
\end{figure}

\subsubsection{Evaluation Protocol}
We evaluate our Global Translation Estimation module on the TotalCapture and IMUPoser datasets, as DIP-IMU lacks translation data. Like prior work~\cite{yi2021transpose,yi2022physical}, we use the Finetune DIP-IMU protocol (Section~\ref{subsubsec:eval_protocol}), that is we train on AMASS and fine-tune on 8 participants of DIP-IMU to track the Root Translation Error (Euclidean norm of the cumulative distance errors within 1 second). 

\subsubsection{Accuracy across Datasets and Body Regions}
On the TotalCapture and IMUPoser dataset, our mean root translation error across all device combinations is 27.55 and 17.63 cm respectively. Interestingly, for both IMUPoser and TotalCapture datasets, we observe only a slight decrease in error when increasing the number of devices from one to two (6.1\%) and no significant improvement (4.0\%) when increasing from two devices to three. Analysing the error across different body regions for the single device scenario (Figure~\ref{fig:translation-error}) (A), we see that a device in the pocket has a much lower error (14.8 cm) compared to that on the wrist (25.7 cm) or the head (29.7 cm). This can be attributed to the legs capturing most of the locomotion data during translation, resulting in marginal gains from sensors on the upper-body. Figure~\ref{fig:translation-error} (B) shows the the cumulative distance error over time. 



\begin{figure}[b]
    \centering
    \includegraphics[width=0.8\columnwidth]{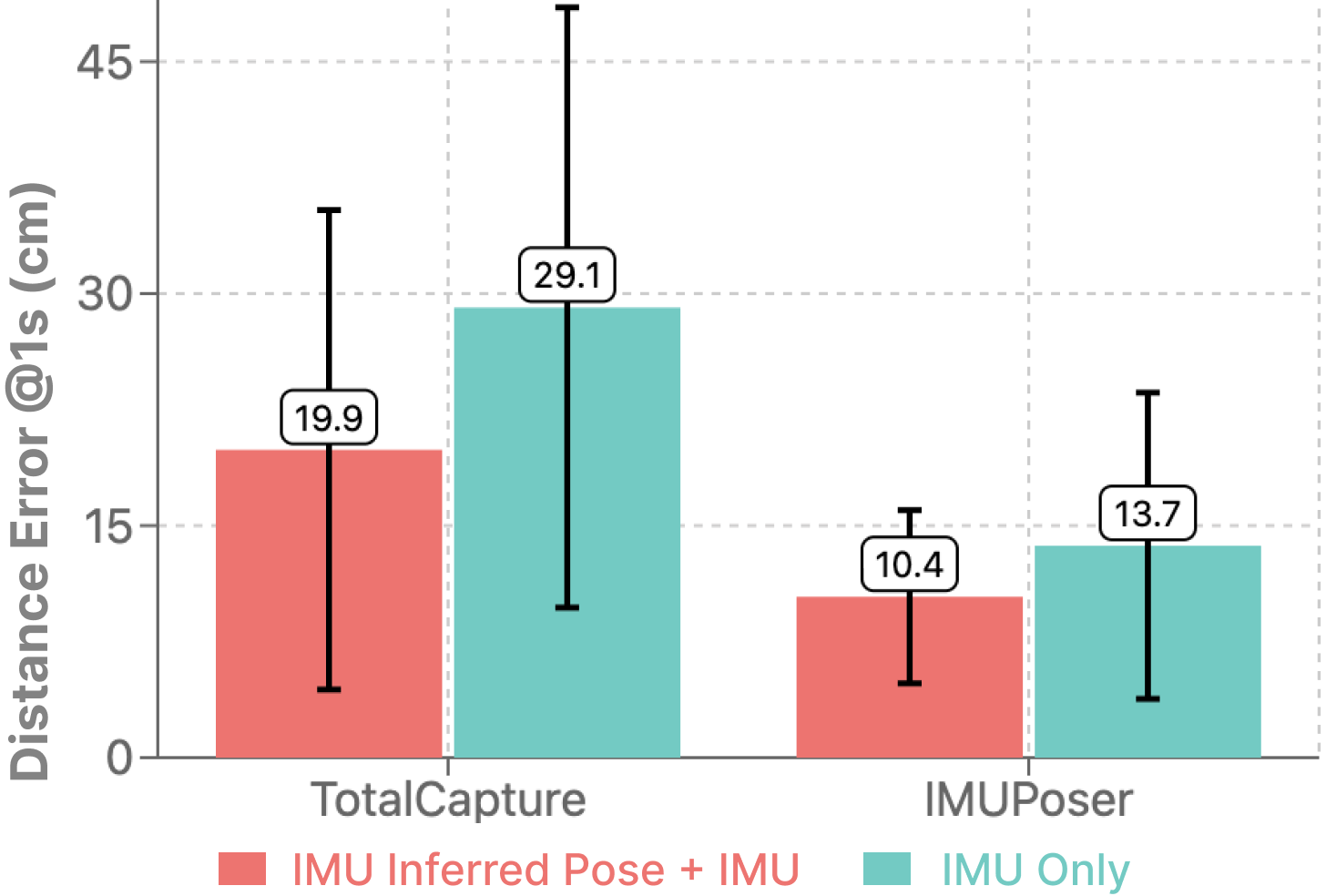}
    \caption{Benefits of using high-order digitization (i.e., IMU inferred poses) for estimating global translation.}
    \label{fig:translation-ablation}
\end{figure}

\subsubsection{Ablation Study}
We perform ablation studies to understand the impact of key components in our system and their effects on performance. At the core of our system lies a subtle yet powerful concept: higher-order digitization (e.g., body pose) improves lower-order digitizations (e.g., steps). To quantify this idea, we run an ablation study of our translation estimation technique using \textit{both} IMU data and the corresponding full-body pose inferred from it versus using only IMU data. Figure~\ref{fig:translation-ablation} summarizes our results. Our IMU-only, direct regression has an error of 21.4 cm across both datasets, while our integrated (IMU + IMU inferred pose) approach decreases error by 29.4\% to 15.1 cm.

Building on the multi-stage architecture, we further evaluate the impact of two additional components: jerk loss and physics refinement. These elements were designed to enhance motion smoothness and physical plausibility. For the IMUPoser dataset, the jerk loss reduces jitter by 23.9\% and translation error by 3.33\%, but increases mean pose error by 0.05\%. Further, the physics-aware refinement reduces jitter by 29.7\% and translation error by 0.4\%, but increases the mean pose error by 0.7\%. The negligible increase in mean pose error is expected, as it may occasionally over-smooth the motion. This phenomenon is also seen in the PIP~\cite{yi2022physical}. We believe that significant improvements in jitter and translation far outweigh the minimal increase in pose error, resulting in a more realistic motion.

\subsubsection{Comparison with prior work}
To the best of our knowledge, no other works have explored both \textit{full-body pose} and \textit{translation} from such a sparse set of commodity IMUs. IMUPoser~\cite{mollyn2023imuposer}, which also targets consumer devices, does not estimate global translation. On the TotalCapture dataset, TransPose (6 IMUs) has a translation error of 12.8 cm while that of MobilePoser is 19.9 cm when a single IMU device is placed in the pocket. Unsurprisingly, a commercial grade, 6 IMU-based system has higher accuracy due to their waist and knee mounted sensors, which capture larger ranges of locomotion compared to devices carried in the pocket.

%% file: sections/5_applications.tex
\begin{figure}
    \centering
    \includegraphics[width=\columnwidth]{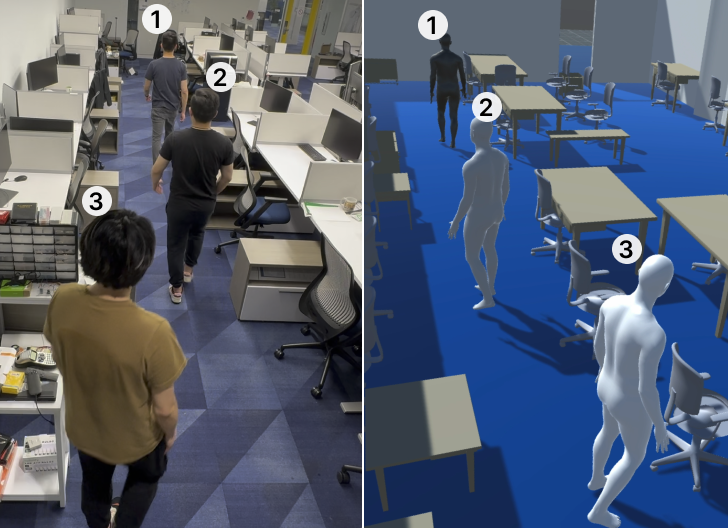}
    \caption{Example indoor navigation application where MobilePoser digitizes multiple users within an office space.}
    \label{fig:indoor-navigation}
\end{figure}

\section{Example Uses}

MobilePoser enables full-body pose estimation with global motion tracking using devices that users already own, opening up a wide range of novel applications. This section showcases three proof-of-concept applications in indoor navigation, gaming, and healthcare to illustrate MobilePoser's unique capabilities and potential impact.

\subsection{Indoor Localization and Navigation}


To demonstrate MobilePoser's potential in this domain, we scan an office space using the PolyCam~\cite{polycam2023} LiDAR scanner app with an Apple iPhone 15 Pro. As shown in Figure~\ref{fig:indoor-navigation}, multiple users walk through the virtual office space, with their interactions and movements seamlessly digitized and represented in real-time. Here, one user has a phone in their pocket and a watch on their wrist, while the other two only have a phone in their pocket. By leveraging the IMUs in these consumer devices, MobilePoser enables accurate indoor navigation and localization without the need for additional infrastructure or specialized hardware. This opens up exciting possibilities for applications such as indoor way finding, context-aware virtual assistants, and immersive virtual tours. 



\subsection{Mobile Gaming Experiences}

\begin{figure}
    \centering
    \includegraphics[width=\columnwidth]{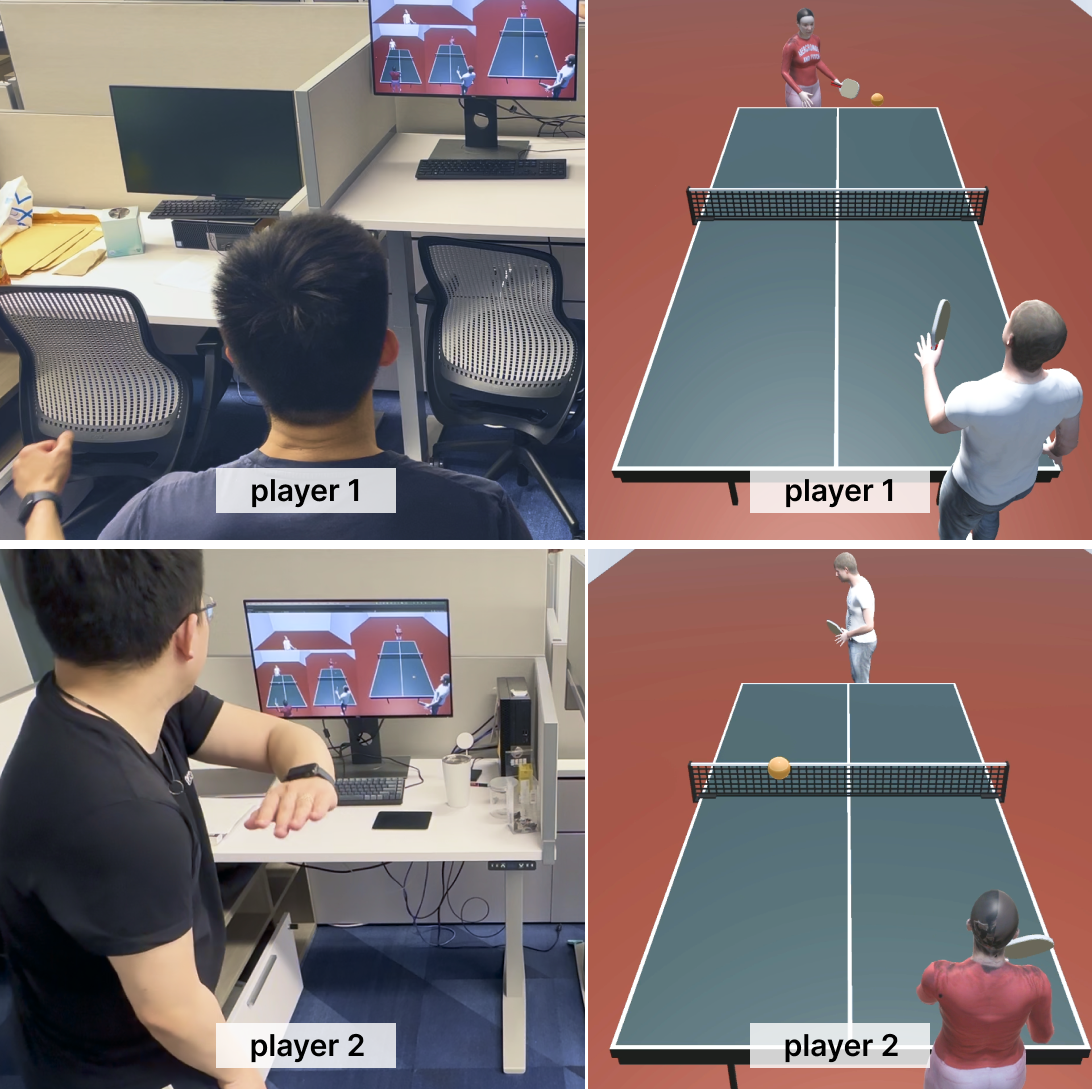}
    \caption{In this table tennis game users can move around the table freely and use their wrist-instrumented hand to control their racket.}
    \label{fig:pingpong}
\end{figure}

To showcase this potential, we developed a virtual table tennis game (Figure~\ref{fig:pingpong}) that allows users to play remotely with others, similar to how Nintendo games are played in front of a TV. Each player has a phone in their pocket and a watch on the dominant (left) hand, which is controlling the racket. Players can freely move within their local space to control their avatars, adding a new level of physical interaction to the gaming experience. MobilePoser's ability to track full-body movements using everyday devices eliminates the need for specialized controllers, making immersive gaming experiences more accessible to a wider audience.

\subsection{Fitness and Wellness}

\begin{figure}
    \centering
    \includegraphics[width=\columnwidth]{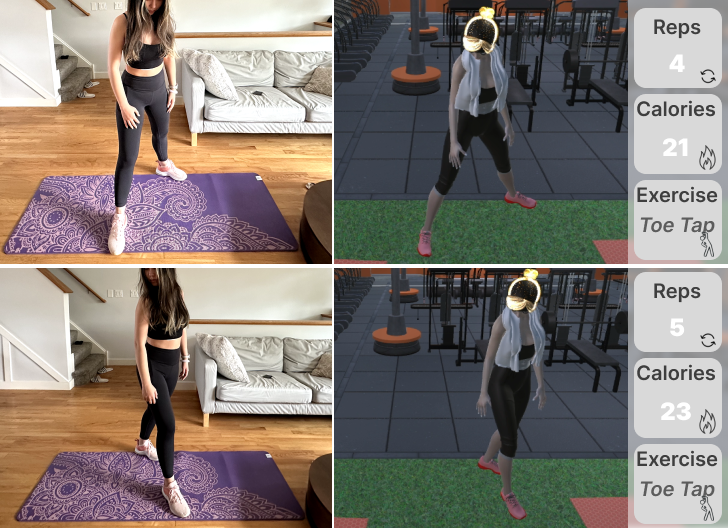}
    \caption{MobilePoser's full-body pose and locomotion can be used to automatically detect and count exercise repetitions, better estimate calories and monitor form. }
    \label{fig:gym}
\end{figure}

MobilePoser has the potential to revolutionize fitness tracking and rehabilitation by providing accurate, real-time feedback on a user's movements and poses without the need for external sensors or camera setups. This enables users to monitor their exercise form, track progress, and receive personalized guidance using the devices they already own. In this example (Figure~\ref{fig:gym}), a user performs a workout routine while MobilePoser captures the session using the IMU data from the smartphone in the user's pocket. This not only allows the user to review their performance and track progress over time but also enables remote monitoring by fitness instructors or physical therapists. Moreover, MobilePoser's ability to track full-body movements facilitates  interactive rehabilitation regimens \cite{ahuja2021pose} and other passive health sensing applications such as gait analysis \cite{nishiguchi2012reliability} or hyperactivity detection \cite{arakawa2023lemurdx}, among others. 

%% file: sections/6_opensource.tex
\section{Open Source}

To enable other researchers and practitioners to build upon our work, we release our pre-trained models, data pre-processing scripts, and model training code as open-source software at: \url{https://github.com/SPICExLAB/MobilePoser}. By making our work fully reproducible and extensible, we hope to accelerate research and development in the field of mobile motion capture using everyday devices. 


%% file: sections/7_limitations_future_work.tex
\section{Limitations and Future Work}

While MobilePoser demonstrates promising results in estimating full-body pose and translation using minimal instrumentation, there are several limitations and opportunities for future work. First, as a purely inertial-based technique, MobilePoser's translation estimation is still susceptible to drift, particularly when devices deviate from their calibrated positions. This can occur when users wear loose clothing, causing the phone in the pocket to move around and resulting in orientation changes. To address this issue, future work could explore re-calibration techniques based on stationary poses or leverage additional sensory information, such as GPS, UWB or visual odometry, to correct for drift.

Second, akin to prior wor, our evaluation has limitations of being tested on lab collected datasets. All the test datasets (DIP, TotalCapture, IMUPoser) were collected in lab settings due to the need for an accurate external ground truth motion capture system. Although we empirically demonstrate that MobilePoser works in real-world settings (as seen in the accompanying video), we acknowledge the need for future datasets captured in-the-wild.


Another limitation of MobilePoser, much like other prior works \cite{yi2021transpose, mollyn2023imuposer,huang2018deep,yi2022physical}, is the need for a calibration step. Currently, users first stand in a T-pose, which aligns the IMU data with the training data based on the SMPL kinematic model. While this calibration process is acceptable for some use cases, such as gaming, it may be less desirable for applications that demand seamless interactions, like indoor navigation. Future work could investigate more natural and unobtrusive calibration procedures, such as detecting common poses like standing with arms by the side using UWB, similar to SmartPoser~\cite{devrio2023smartposer}.


In conclusion, while MobilePoser presents a significant step forward in enabling full-body pose and translation estimation using everyday devices, there remain several avenues for future research to extend the capabilities of this approach.

%% file: sections/8_conclusion.tex
\section{Conclusion}

In this paper, we present MobilePoser, a real-time, on-device system for estimating full-body pose and translation using IMUs in consumer mobile devices (phones, watches, earbuds). By leveraging a multi-stage approach that combines data-driven learning and physics-based optimization, MobilePoser achieves state-of-the-art accuracy while remaining lightweight and efficient. Our extensive evaluation on public datasets demonstrates clear improvements over prior work, both in terms of full-body pose estimation accuracy and enabling novel global translation estimation. Furthermore, we showcase the potential of MobilePoser through a series of proof-of-concept applications in gaming, fitness, and indoor navigation, highlighting its ability to enable new and immersive experiences using the devices people already own.